\documentclass[aps,prd,superscriptaddress,twoside,twocolumn,nofootinbib,showpacs,floatfix]{revtex4-1}
\usepackage{amsmath,amssymb}
\usepackage{graphicx}
\usepackage{subfigure}
\usepackage{color}
\usepackage{ulem}
\pdfpagewidth=\paperwidth
\pdfpageheight=\paperheight
\allowdisplaybreaks
\newcommand*{\slashed}[1]{{#1\!\!\!/}}
\newcommand*{\hc}{\text{H.\,c.}}

\begin{document}

\title{\boldmath Nucleon and $\Delta$ resonances in  $\gamma p \to K^+ \Sigma^0(1385)$ photoproduction}

\author{Ai-Chao Wang}
\affiliation{School of Nuclear Science and Technology, University of Chinese Academy of Sciences, Beijing 100049, China}

\author{Wen-Ling Wang}
\affiliation{School of Physics, Beihang University, Beijing 100191, China}

\author{Fei Huang}
\email[Corresponding author. Email: ]{huangfei@ucas.ac.cn}
\affiliation{School of Nuclear Science and Technology, University of Chinese Academy of Sciences, Beijing 100049, China}

\date{\today}

\begin{abstract}
The photoproduction of $\gamma p \to K^+ \Sigma^0(1385)$ is investigated based on an effective Lagrangian approach using the tree-level Born approximation, with the purpose of understanding the reaction mechanisms and resonance contents and their associated parameters in this reaction. In addition to the $t$-channel $K$ and $K^\ast(892)$ exchanges, $s$-channel nucleon ($N$) exchange, $u$-channel $\Lambda$ exchange, and  generalized contact term, the exchanges of a minimum number of $N$ and $\Delta$ resonances in the $s$ channel are taken into account in constructing the reaction amplitudes to describe the experimental data. It is found that the most recent differential cross-section data from the CLAS Collaboration can be well reproduced by including one of the $N(1895){1/2}^-$, $\Delta(1900){1/2}^-$, and $\Delta(1930){5/2}^-$ resonances. The reaction mechanisms of $\gamma p \to K^+ \Sigma^0(1385)$ are discussed in detail, and the predictions of the beam and target asymmetries for this reaction are given. The cross sections of $\gamma p \to K^0 \Sigma^+(1385)$ are shown to be able to further constrain the theoretical models and pin down the resonance contents for $\gamma p \to K^+ \Sigma^0(1385)$.
\end{abstract}



\maketitle

\section{Introduction}   \label{Sec:intro}

The study of nucleon resonances ($N^\ast$'s) and $\Delta$ resonances ($\Delta^\ast$'s) has always been of great interest in hadron physics, since a deeper understanding of $N$ and $\Delta$ resonances is essential to get insight into the nonperturbative regime of quantum chromodynamics. It is known that most of our current knowledge about $N^\ast$'s and $\Delta^\ast$'s mainly comes from $\pi N$ scattering or $\pi$ photoproduction reactions. Nevertheless, quark models \cite{Isgur:1978,Capstick:1986,Loring:2001} predicated much more $N^\ast$'s and $\Delta^\ast$'s than experimentally observed. One possible explanation of this situation is that some of the $N^\ast$'s and $\Delta^\ast$'s couple weakly to $\pi N$ but strongly to other meson production reactions. Therefore, it is interesting and necessary to study the $N^\ast$'s and $\Delta^\ast$'s in production reactions of mesons other than $\pi$. In the present work, we concentrate on the photoproduction of $K^+\Sigma^0(1385)$. Since the threshold of $K\Sigma(1385)$ is much higher than that of $\pi N$, the $K\Sigma(1385)$ photoproduction reaction is rather suitable to investigate the $N^\ast$'s and $\Delta^\ast$'s in the less-explored higher-energy region.

Experimentally, in the 1970s there was limited experimental data with large error bars on the total cross sections for $\gamma p \to K^+ \Sigma^0(1385)$ \cite{CBCG67,DBCG67,ABBH69}. In 2013, differential cross-section data for $\gamma p \to K^+ \Sigma^0(1385)$ became available in the center-of-mass energy range $W\approx 2.0- 2.8$ GeV from the CLAS Collaboration at the Thomas Jefferson National Accelerator Facility \cite{Mori:2013}. These new differential cross-section data provided stronger constrains on the theoretical amplitudes for $\gamma p \to K^+ \Sigma^0(1385)$; however, they are scarce at very backward and very forward angles, which leads to high uncertainties in the theoretical investigations of this reaction.

Theoretically, based on an effective Lagrangian approach, a hadronic model for $\gamma p \to K^+ \Sigma^0(1385)$ was proposed in 2008 in Ref.~\cite{Yong:2008}, where eight $N$ and $\Delta$ resonances around $2$ GeV (among tens of resonances predicated by a quark model \cite{Capstick:1998}) were considered. In this pioneering work, the resonance masses and resonance hadronic and electromagnetic couplings were taken to have the corresponding values calculated in the quark model \cite{Capstick:1998}, and the resonance widths were set to a common value of $300$ MeV. It was found that the resonance contributions mainly come from the $\Delta(2000)5/2^+$, $\Delta(1940)3/2^-$, $N(2120)3/2^-$ [previously called $N(2080)3/2^-$], and $N(2095)3/2^-$ resonances. One notices that in this work, although the calculated total cross sections are in good agreement with the corresponding preliminary data, there are still some discrepancies between their predicated differential cross sections with the CLAS data published in $2013$ \cite{Mori:2013}, especially in the near-threshold energy region. In $2014$, the reaction $\gamma p \to K^+ \Sigma^0(1385)$ was investigated within a Regge-plus-resonance approach in Ref.~\cite{hejun:2014}. The theoretical framework employed in this work is similar to that proposed in Ref.~\cite{Yong:2008}, with the major differences being the following: (i) in Ref.~\cite{hejun:2014} the $t$-channel $K$ and $K^\ast(892)$ exchanges were considered in a particular Regge type instead of a pure Feynman type, which introduced four additional parameters in the weighting function (form factors), (ii) nine instead of eight $N$ and $\Delta$ resonances around $2$ GeV (among tens of resonances predicated by the quark model of Ref.~\cite{Capstick:1998}) were considered, and (iii) a common width of $500$ MeV (instead of $300$ MeV) was used for all resonances. In Ref.~\cite{hejun:2014}, the CLAS differential cross-section data \cite{Mori:2013} were well reproduced, and it was found that the cross sections of $\gamma p \to K^+ \Sigma^0(1385)$ are dominated by the contact term, while the contributions from all of the considered resonances are much smaller than those in Ref.~\cite{Yong:2008} due to the much larger resonance width. In $2017$, the reaction $\gamma p \to K^+ \Sigma^0(1385)$ was studied in a Regge model in Ref.~\cite{Yu:2017}, where the Reggeized $t$-channel $K$, $K^\ast(892)$, and $K_2^\ast(1430)$ exchanges were considered, and it was found that the reaction mechanism is featured by the dominance of the contact term plus the $K$ exchange with the role of the $K_2^\ast(1430)$ following rather than the $K^\ast(892)$. In Ref.~\cite{Yu:2017} the total cross-section data were well reproduced, but considerable discrepancies were still seen in the calculated differential cross sections compared with the corresponding data due to the lack of $N$ and $\Delta$ resonances.

In the present work, we investigate the $\gamma p \to K^+ \Sigma^0(1385)$ reaction within an effective Lagrangian approach using the tree-level Born approximation. In addition to the $t$-channel $K$ and $K^\ast(892)$ exchanges, $s$-channel $N$ exchange, $u$-channel $\Lambda$ exchange, and generalized contact term, we consider as few as possible $N$ and $\Delta$ resonances in the $s$ channel to describe the most recent differential cross-section data from the CLAS Collaboration \cite{Mori:2013}. The $t$-, $s$-, and $u$-channel amplitudes are obtained by evaluating the corresponding Feynman diagrams, and the generalized contact term is constructed to ensure the gauge invariance of the full photoproduction amplitudes. With regard to the $N$ and $\Delta$ resonances, the present work differs considerably from Refs.~\cite{Yong:2008,hejun:2014} in the following three respects: (i) in the present work we introduce as few $N$ and $\Delta$ resonances as possible to reproduce the data, while in Refs.~\cite{Yong:2008,hejun:2014} eight or nine resonances (among tens of resonances predicated by a quark model calculation \cite{Capstick:1998}) were considered; (ii) in the present work the masses and widths of the resonances are fixed to the values advocated by the Particle Data Group (PDG) \cite{Tanabashi:2018}, while in Refs.~\cite{Yong:2008,hejun:2014} the masses of the resonances were taken from a quark model calculation \cite{Capstick:1998} and the widths for all of the resonances were set to a common value of $300$ or $500$ MeV, respectively; (iii) in the present work the resonance couplings are treated as parameters to be determined by fits to the data, while in Refs.~\cite{Yong:2008,hejun:2014} they were fixed by the decay amplitudes calculated in a quark model  \cite{Capstick:1998}. We believe that such an independent analysis of the available data for $\gamma p \to K^+ \Sigma^0(1385)$ as performed in the present work is necessary and useful for a better understanding of the reaction mechanisms, resonance contents, and associated resonance parameters in this reaction.

The present paper is organized as follows. In Sec.~\ref{Sec:formalism}, we briefly introduce the framework of our theoretical model, including the generalized contact current, effective interaction Lagrangians, resonance propagators, and phenomenological form factors employed in the present work. In Sec.~\ref{Sec:results}, we present our theoretical results for the differential and total cross sections for $\gamma p \to K^+ \Sigma^0(1385)$, and discuss the contributions from individual terms. Furthermore, the beam and target asymmetries for $\gamma p \to K^+ \Sigma^0(1385)$ and the total cross sections for $\gamma p \to K^0 \Sigma^+(1385)$ are shown and discussed in this section. Finally, a brief summary and conclusions are given in Sec.~\ref{sec:summary}.

\section{Formalism}  \label{Sec:formalism}

\begin{figure}[tbp]
\subfigure[~$s$ channel]{
\includegraphics[width=0.45\columnwidth]{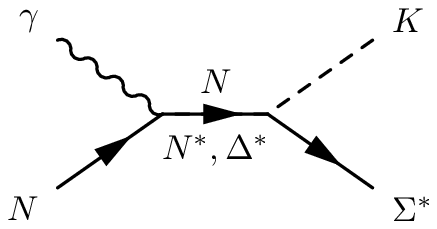}}  {\hglue 0.4cm}
\subfigure[~$t$ channel]{
\includegraphics[width=0.45\columnwidth]{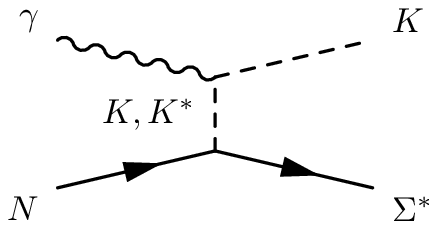}} \\[6pt]
\subfigure[~$u$ channel]{
\includegraphics[width=0.45\columnwidth]{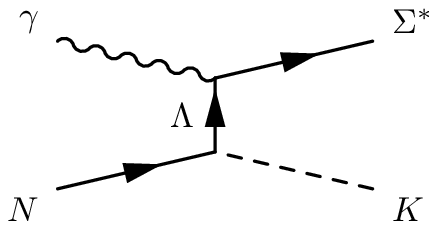}} {\hglue 0.4cm}
\subfigure[~Interaction current]{
\includegraphics[width=0.45\columnwidth]{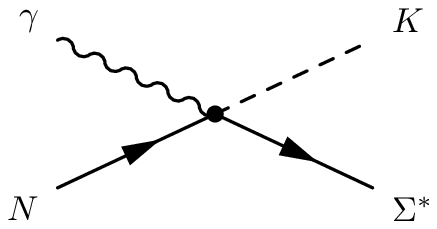}}
\caption{Generic structure of the amplitude for $\gamma p\to K^+ \Sigma^0(1385)$. Time proceeds from left to right. The symbols $\Sigma^\ast$ and $K^\ast$ denote $\Sigma(1385)$ and $K^\ast(892)$, respectively.}
\label{FIG:feymans}
\end{figure}

Following the full field-theoretical approach of Refs.~\cite{Haberzettl:1997,Haberzettl:2006}, the full photoproduction amplitudes for $\gamma N \to K \Sigma(1385)$ can be expressed as
\begin{equation}
M^{\nu\mu} = M^{\nu\mu}_s + M^{\nu\mu}_t + M^{\nu\mu}_u + M^{\nu\mu}_{\rm int},  \label{eq:amplitude}
\end{equation}
with $\nu$ and $\mu$ being the Lorentz indices of $\Sigma(1385)$ and the photon $\gamma$, respectively. The first three terms $M^{\nu\mu}_s$, $M^{\nu\mu}_t$, and $M^{\nu\mu}_u$ stand for the $s$-, $t$-, and $u$-channel pole diagrams, respectively, with $s$, $t$, and $u$ being the Mandelstam variables of the internally exchanged particles. They arise from the photon attaching to the external particles in the underlying $KN\Sigma(1385)$ interaction vertex. The last term, $M^{\nu\mu}_{\rm int}$, stands for the interaction current that arises from the photon attaching to the internal structure of the $KN\Sigma(1385)$ interaction vertex. All four terms in Eq.~(\ref{eq:amplitude}) are diagrammatically depicted in Fig.~\ref{FIG:feymans}.

In the present work, the following contributions (as shown in Fig.~\ref{FIG:feymans}) are considered in constructing the $s$-, $t$-, and $u$-channel amplitudes: (i) $N$, $N^\ast$, and $\Delta^\ast$ exchanges in the $s$ channel, (ii) $K$ and $K^\ast(892)$ exchanges in the $t$ channel, and (iii) $\Lambda$ hyperon exchange in the $u$ channel. We mention that, following Refs.~\cite{Yong:2008,hejun:2014}, the exchanges of other hyperon states in the $u$ channel are omitted in the present work. Using an effective Lagrangian approach, one can, in principle, obtain explicit expressions for these amplitudes by evaluating the corresponding Feynman diagrams. However, the exact calculation of the interaction current $M^{\nu\mu}_{\rm int}$ is impractical, as it obeys a highly nonlinear equation and contains diagrams with very complicated interaction dynamics. Furthermore, the introduction of phenomenological form factors makes it impossible to calculate the interaction current exactly even in principle. Following Refs.~\cite{Haberzettl:1997,Haberzettl:2006,Huang:2012,Huang:2013}, we model the interaction current by a generalized contact current, which effectively accounts for the interaction current arising from the unknown parts of the underlying microscopic model,
\begin{equation}
M^{\nu\mu}_{\rm int} = \Gamma^\nu_{\Sigma^\ast N K}(q) C^\mu + M^{\nu\mu}_{\rm KR} f_t.  \label{eq:Mint}
\end{equation}
Here $\nu$ and $\mu$ are Lorentz indices for $\Sigma(1385)$ and $\gamma$, respectively; $\Gamma^\nu_{\Sigma^\ast NK}(q)$ is the vertex function of $\Sigma(1385) NK$ coupling given by the Lagrangian of Eq.~(\ref{eq:L_KNSt}),
\begin{equation}
\Gamma^\nu_{\Sigma^\ast NK}(q) = - \frac{g_{\Sigma^\ast NK}}{M_K} q^\nu,
\end{equation}
with $q$ being the four-momentum of the outgoing $K$ meson; $M_{\rm KR}^{\nu\mu}$ is the Kroll-Ruderman term given by the Lagrangian of Eq.~(\ref{eq:L_rKNSt}),
\begin{equation}
M^{\nu\mu}_{\rm KR} = \frac{g_{\Sigma^\ast NK}}{M_K}  g^{\nu\mu} TQ_K,
\end{equation}
with $T$ denoting the isospin factor of the $\Sigma(1385) NK$ coupling and $Q_K$ being the electric charge of the outgoing $K$ meson; $f_t$ is the phenomenological form factor attached to the amplitude of $t$-channel $K$ exchange, which is given in Eq.~(\ref{eq:ff_M}); $C^\mu$ is an auxiliary current, which is nonsingular and is introduced to ensure that the full photoproduction amplitudes of Eq.~(\ref{eq:amplitude}) are fully gauge invariant. Following Refs.~\cite{Haberzettl:2006,Huang:2012}, we choose $C^\mu$ for $\gamma p \to K^+ \Sigma^{0}(1385)$ as
\begin{equation}   \label{eq:Cmu}
C^\mu =  - Q_{K} \frac{f_t-\hat{F}}{t-q^2}  (2q-k)^\mu - Q_N \frac{f_s-\hat{F}}{s-p^2} (2p+k)^\mu,
\end{equation}
with
\begin{equation} \label{eq:Fhat-Kstp}
\hat{F} = 1 - \hat{h} \left(1 -  f_s\right) \left(1 - f_t\right).
\end{equation}
Here $p$, $q$, and $k$ are the four-momenta for the incoming $N$, outgoing $K$, and incoming photon, respectively; $Q_{N\left(K\right)}$ is the electric charge of $N\left(K\right)$; $f_s$ and $f_t$ are the phenomenological form factors for $s$-channel $N$ exchange and $t$-channel $K$ exchange, respectively; $\hat{h}$ is an arbitrary function that goes to unity in the high-energy limit and is set to be $\hat{h}=1$ in the present work for simplicity.

In the rest of this section,  we present the effective Lagrangians, resonance propagators, and phenomenological form factors employed in the present work.

\subsection{Effective Lagrangians} \label{Sec:Lagrangians}

The effective interaction Lagrangians used in the present work for the production amplitudes are given below. For further convenience, we define the operators
\begin{equation}
\Gamma^{(+)}=\gamma_5  \quad  \text{and} \quad  \Gamma^{(-)}=1,
\end{equation}
the field
\begin{equation}
\Sigma^\ast = \Sigma(1385),
\end{equation}
and the field-strength tensors
\begin{equation}
F^{\mu\nu} = \partial^{\mu}A^\nu-\partial^{\nu}A^\mu,
\end{equation}
with $A^\mu$ denoting the electromagnetic field.

The electromagnetic interaction Lagrangians required to calculate the nonresonant Feynman diagrams are
\begin{eqnarray}
\mathcal{L}_{\gamma KK} &=& ie \left[ K^+ \left(  \partial^\mu K^- \right)  -  K^- \left(\partial^\mu K^+ \right) \right] A_\mu,   \\[6pt]
{\cal L}_{\gamma K{K^\ast}} &=& e\frac{g_{\gamma K{K^\ast}}}{M_K}\varepsilon^{\alpha \mu \lambda \nu}\left(\partial_\alpha A_\mu\right)\left(\partial_\lambda K\right)K^\ast_\nu, \label{Lag:gKKst} \\[6pt]
{\cal L}_{NN\gamma} &=& -\,e \bar{N} \left[ \left( \hat{e} \gamma^\mu - \frac{ \hat{\kappa}_N} {2M_N}\sigma^{\mu \nu}\partial_\nu\right) A_\mu\right] N, \\[6pt]
{\cal L}_{\Sigma^\ast \Lambda \gamma} &=& -ie\frac{g^{(1)}_{\Sigma^\ast \Lambda \gamma}}{2M_N} \bar{\Sigma}^\ast_\mu \gamma_\nu \gamma_5 F^{\mu \nu} \Lambda \nonumber \\
&& +\,e\frac{g^{(2)}_{\Sigma^\ast \Lambda \gamma}}{\left(2M_N\right)^2} \bar{\Sigma}^\ast_\mu \gamma_5 F^{\mu \nu}\partial_\nu \Lambda  + \hc,
\end{eqnarray}
where $e$ is the elementary charge unit and $\hat{e}$ stands for the charge operator; $\hat{\kappa}_N = \kappa_p\left(1+\tau_3\right)/2 + \kappa_n\left(1-\tau_3\right)/2$, with the anomalous magnetic moments $\kappa_p=1.793$ and $\kappa_n=-1.913$; $M_N$ and $M_K$ stand for the masses of $N$ and $K$, respectively; $\varepsilon^{\alpha \mu \lambda \nu}$ is the totally antisymmetric Levi-Civita tensor with $\varepsilon^{0123}=1$. The value of the electromagnetic coupling $g_{\gamma K K^\ast}$ is determined by fitting the radiative decay width of $K^\ast(892) \to K\gamma$ given by the PDG \cite{Tanabashi:2018}, which leads to $g_{\gamma K^\pm K^{\ast\pm}}=0.413$ with the sign inferred from $g_{\gamma \pi \rho}$ \cite{Garcilazo:1993} via the flavor SU(3) symmetry considerations in conjunction with the vector-meson dominance assumption. The coupling constants $g^{(1)}_{\Sigma^\ast \Lambda \gamma}$ and $g^{(2)}_{\Sigma^\ast \Lambda \gamma}$ are constrained by the radiative decay width of $\Gamma_{\Sigma^0(1385)\to\Lambda \gamma} = 0.45$ MeV \cite{Tanabashi:2018}, and thus only one of them is free. In the present work, we treat the ratio $g^{(1)}_{\Sigma^\ast \Lambda \gamma}/g^{(2)}_{\Sigma^\ast \Lambda \gamma}$ as a fit parameter.

The resonance-nucleon-photon transition Lagrangians are
\begin{eqnarray}
{\cal L}_{RN\gamma}^{1/2\pm} &=& e\frac{g_{RN\gamma}^{(1)}}{2M_N}\bar{R} \Gamma^{(\mp)}\sigma_{\mu\nu} \left(\partial^\nu A^\mu \right) N  + \hc, \\[6pt]
{\cal L}_{RN\gamma}^{3/2\pm} &=& -\, ie\frac{g_{RN\gamma}^{(1)}}{2M_N}\bar{R}_\mu \gamma_\nu \Gamma^{(\pm)}F^{\mu\nu}N \nonumber \\
&& +\, e\frac{g_{RN\gamma}^{(2)}}{\left(2M_N\right)^2}\bar{R}_\mu \Gamma^{(\pm)}F^{\mu \nu}\partial_\nu N + \hc, \\[6pt]
{\cal L}_{RN\gamma}^{5/2\pm} & = & e\frac{g_{RN\gamma}^{(1)}}{\left(2M_N\right)^2}\bar{R}_{\mu \alpha}\gamma_\nu \Gamma^{(\mp)}\left(\partial^{\alpha} F^{\mu \nu}\right)N \nonumber \\
&& \pm\, ie\frac{g_{RN\gamma}^{(2)}}{\left(2M_N\right)^3}\bar{R}_{\mu \alpha} \Gamma^{(\mp)}\left(\partial^\alpha F^{\mu \nu}\right)\partial_\nu N \nonumber \\
&& +\, \hc, \\[6pt]
{\cal L}_{RN\gamma}^{7/2\pm} &=&  ie\frac{g_{RN\gamma}^{(1)}}{\left(2M_N\right)^3}\bar{R}_{\mu \alpha \beta}\gamma_\nu \Gamma^{(\pm)}\left(\partial^{\alpha}\partial^{\beta} F^{\mu \nu}\right)N \nonumber \\
&& -\, e\frac{g_{RN\gamma}^{(2)}}{\left(2M_N\right)^4}\bar{R}_{\mu \alpha \beta} \Gamma^{(\pm)} \left(\partial^\alpha \partial^\beta F^{\mu \nu}\right) \partial_\nu N  \nonumber \\
&&  +\, \hc,
\end{eqnarray}
where $R$ designates the $N$ or $\Delta$ resonance, and the superscript of ${\cal L}_{RN\gamma}$ denotes the spin and parity of the resonance $R$. The coupling constants $g_{RN\gamma}^{(i)}$ $(i=1,2)$ can, in principle, be determined by the resonance radiative decay amplitudes. Nevertheless, since the resonance hadronic coupling constants are unknown due to the lack of experimental information on the resonance decay to $K\Sigma(1385)$, we treat the products of the electromagnetic and hadronic coupling constants---which are relevant to the production amplitudes---as fit parameters in the present work.

The effective Lagrangians for meson-baryon interactions are
\begin{eqnarray}
\mathcal{L}_{\Sigma^\ast NK} &=& \frac{g_{\Sigma^\ast NK}}{M_K} \bar{\Sigma}^{\ast \mu} \left( \partial_\mu \bar{K} \right) N  + \hc,    \label{eq:L_KNSt}    \\[6pt]
{\cal L}_{\Lambda NK} &=& -\, \frac{g_{\Lambda NK}}{2M_N}\bar{\Lambda} \gamma_5 \gamma^\mu \left( \partial_\mu K \right) N + \hc,  \label{eq:L_LNK}    \\[6pt]
{\cal L}_{\Sigma^\ast N{K^\ast}} &=& -\, i\frac{g_{\Sigma^\ast N K^\ast}^{(1)}}{2M_{N}}{\bar\Sigma}^\ast_\mu \gamma_\nu \gamma_5 {K^\ast}^{\mu \nu}N \nonumber \\
&& +\, \frac{g_{\Sigma^\ast N K^\ast}^{(2)}}{\left(2M_{N}\right)^2}{\bar\Sigma}^\ast_\mu \gamma_5 {K^\ast}^{\mu \nu}\partial_\nu N \nonumber \\
&& -\, \frac{g_{\Sigma^\ast N K^\ast}^{(3)}}{\left(2M_{N}\right)^2}{\bar\Sigma}^\ast_\mu \gamma_5\left(\partial_\nu {K^{\ast\mu \nu}}\right) N + \hc.
\end{eqnarray}
The coupling constant $g_{\Lambda NK} \approx -14$ is determined by the flavor SU(3) symmetry,
\begin{eqnarray}
g_{\Lambda NK} = -\frac{3\sqrt{3}}{5} g_{NN\pi},   \label{g_LamNK}
\end{eqnarray}
with $g_{NN\pi} = 13.46$. The coupling constants $g_{\Sigma^\ast NK}$ and $g^{(1)}_{\Sigma^\ast NK^\ast}$ are also fixed by the flavor SU(3) symmetry \cite{Swart:1963,Ronchen:2013},
\begin{eqnarray}
\frac{g_{\Sigma^\ast NK}}{M_K} &=& -\, \frac{1}{\sqrt{6}}\frac{g_{\Delta N\pi }}{M_\pi},  \label{g_SigstNK}    \\[6pt]
g^{(1)}_{\Sigma^\ast NK^\ast} &=& -\,\frac{1}{\sqrt{6}} g_{\Delta N\rho}.   \label{g_SigstNKst}
\end{eqnarray}
By using the value $g_{\Delta N\pi}=2.23$ determined from the $\Delta$ resonance decay width, $\Gamma_{\Delta\to N\pi}=120$ MeV, and the empirical value $g_{\Delta N\rho}=-39.1$, one gets $g_{\Sigma^\ast NK}=-3.22$ and $g^{(1)}_{\Sigma^\ast NK^\ast}=15.96$. As the $g^{(2)}$ and $g^{(3)}$ terms in the $\Delta N\rho$ interactions have never been seriously studied in the literature, the corresponding coupling constants in $\Sigma^\ast NK^\ast$ vertices, i.e. $g_{\Sigma^\ast N K^\ast}^{(2)}$ and $g_{\Sigma^\ast N K^\ast}^{(3)}$, cannot be determined via flavor SU(3) symmetry, and we ignore these two terms in the present work, following Refs.~\cite{Kim:2011,Kim:2014,Wang:2017,Wang:2018}.

The effective Lagrangians for resonance hadronic vertices can be written as
\begin{eqnarray}
\mathcal{L}_{R\Sigma^\ast K}^{1/2\pm} &=& \frac{g_{R\Sigma^\ast K}^{(1)}}{M_K} \bar{\Sigma}^{\ast}_\mu \Gamma^{(\mp)} \left(\partial^\mu K\right) R + \hc,   \\[6pt]
\mathcal{L}_{R\Sigma^\ast K}^{3/2\pm} &=& \frac{g_{R\Sigma^\ast K}^{(1)}}{M_K} \bar{\Sigma}^{\ast}_\mu  \gamma^\alpha \Gamma^{(\pm)} \left(\partial_\alpha K\right) R^\mu \nonumber \\
&& + \, i \frac{g_{R\Sigma^\ast K}^{(2)}}{M_K^2} \bar{\Sigma}^\ast_\alpha \Gamma^{(\pm)} \left(\partial^\mu \partial^\alpha K\right)  R_\mu + \hc,   \\[6pt]
\mathcal{L}_{R\Sigma^\ast K}^{5/2\pm} &=& i \frac{ g_{R\Sigma^\ast K}^{(1)} }{M_K^2} \bar{\Sigma}^{\ast}_\alpha \gamma^\mu \Gamma^{(\mp)} \left(\partial_\mu \partial_\beta K\right)  R^{\alpha\beta}  \nonumber \\
&& - \, \frac{ g_{R\Sigma^\ast K}^{(2)} }{M_K^3} \bar{\Sigma}^\ast_\mu \Gamma^{(\mp)} \left( \partial^\mu \partial^\alpha \partial^\beta K \right)  R_{\alpha\beta} \nonumber \\
&& + \, \hc,   \\[6pt]
\mathcal{L}_{R\Sigma^\ast K}^{7/2\pm} &=& - \frac{ g_{R\Sigma^\ast K}^{(1)} }{M_K^3} \bar{\Sigma}^{\ast}_\alpha \gamma^\mu \Gamma^{(\pm)} \left(\partial_\mu \partial_\beta \partial_\lambda K\right)  R^{\alpha\beta\lambda}  \nonumber \\
&& - \, i \frac{ g_{R\Sigma^\ast K}^{(2)} }{M_K^4} \bar{\Sigma}^\ast_\mu \Gamma^{(\pm)} \left( \partial^\mu \partial^\alpha \partial^\beta \partial^\lambda K \right)  R_{\alpha\beta\lambda} \nonumber \\
&& + \, \hc.
\end{eqnarray}
In the present work, the $g_{R\Sigma^\ast K}^{(2)}$ terms in ${\cal L}_{R\Sigma^\ast K}^{3/2\pm}$, ${\cal L}_{R\Sigma^\ast K}^{5/2\pm}$, and ${\cal L}_{R\Sigma^\ast K}^{7/2\pm}$ are ignored for the sake of simplicity. The coupling constants $g_{R\Sigma^\ast K}^{(1)}$ are treated as fit parameters. Actually, only the products of the electromagnetic couplings and the hadronic couplings of $N$ or $\Delta$ resonances are relevant to the reaction amplitudes, and these products are what we really fit in practice.

The effective Lagrangian for the Kroll-Ruderman term of $\gamma N \to K\Sigma(1385)$ reads
\begin{equation}
{\cal L}_{\gamma \Sigma^\ast N K} = -iQ_K \frac{g_{\Sigma^\ast NK}}{M_K} \bar{\Sigma}^{\ast \mu} A_\mu \bar{K} N  + \hc,   \label{eq:L_rKNSt}
\end{equation}
which is obtained by the minimal gauge substitution $\partial_\mu \to {\cal D}_\mu\equiv \partial_\mu - i Q_K A_\mu$ in the ${\cal L}_{\Sigma^\ast NK}$ interaction Lagrangian of Eq.~(\ref{eq:L_KNSt}). The coupling constant $g_{\Sigma^\ast NK}$ has been given in Eq.~(\ref{g_SigstNK}).

\subsection{Resonance propagators}

For the spin-$1/2$ resonance propagator, we use the ansatz
\begin{equation}
S_{1/2}(p) = \frac{i}{\slashed{p} - M_R + i \Gamma_R/2},
\end{equation}
where $M_R$ and $\Gamma_R$ are, respectively, the mass and width of the resonance $R$, and $p$ is the resonance four-momentum.

Following Refs.~\cite{Behrends:1957,Fronsdal:1958,Zhu:1999}, the prescriptions of the propagators for resonances with spin $3/2$, $5/2$, and $7/2$ are
\begin{eqnarray}
S_{3/2}(p) &=&  \frac{i}{\slashed{p} - M_R + i \Gamma_R/2} \left( \tilde{g}_{\mu \nu} + \frac{1}{3} \tilde{\gamma}_\mu \tilde{\gamma}_\nu \right),  \\[6pt]
S_{5/2}(p) &=&  \frac{i}{\slashed{p} - M_R + i \Gamma_R/2} \,\bigg[ \, \frac{1}{2} \big(\tilde{g}_{\mu \alpha} \tilde{g}_{\nu \beta} + \tilde{g}_{\mu \beta} \tilde{g}_{\nu \alpha} \big)  \nonumber \\
&& -\, \frac{1}{5}\tilde{g}_{\mu \nu}\tilde{g}_{\alpha \beta}  + \frac{1}{10} \big(\tilde{g}_{\mu \alpha}\tilde{\gamma}_{\nu} \tilde{\gamma}_{\beta} + \tilde{g}_{\mu \beta}\tilde{\gamma}_{\nu} \tilde{\gamma}_{\alpha}  \nonumber \\
&& +\, \tilde{g}_{\nu \alpha}\tilde{\gamma}_{\mu} \tilde{\gamma}_{\beta} +\tilde{g}_{\nu \beta}\tilde{\gamma}_{\mu} \tilde{\gamma}_{\alpha} \big) \bigg],   \\[6pt]
S_{7/2}(p) &=&  \frac{i}{\slashed{p} - M_R + i \Gamma_R/2} \, \frac{1}{36}\sum_{P_{\mu} P_{\nu}} \bigg( \tilde{g}_{\mu_1 \nu_1}\tilde{g}_{\mu_2 \nu_2}\tilde{g}_{\mu_3 \nu_3} \nonumber \\
&& -\, \frac{3}{7}\tilde{g}_{\mu_1 \mu_2}\tilde{g}_{\nu_1 \nu_2}\tilde{g}_{\mu_3 \nu_3} + \frac{3}{7}\tilde{\gamma}_{\mu_1} \tilde{\gamma}_{\nu_1} \tilde{g}_{\mu_2 \nu_2}\tilde{g}_{\mu_3 \nu_3} \nonumber \\
&& -\, \frac{3}{35}\tilde{\gamma}_{\mu_1} \tilde{\gamma}_{\nu_1} \tilde{g}_{\mu_2 \mu_3}\tilde{g}_{\nu_2 \nu_3} \bigg),  \label{propagator-7hf}
\end{eqnarray}
where
\begin{eqnarray}
\tilde{g}_{\mu \nu} &=& -\, g_{\mu \nu} + \frac{p_{\mu} p_{\nu}}{M_R^2}, \\[6pt]
\tilde{\gamma}_{\mu} &=& \gamma^{\nu} \tilde{g}_{\nu \mu} = -\gamma_{\mu} + \frac{p_{\mu}\slashed{p}}{M_R^2},
\end{eqnarray}
and the summation over $P_{\mu}(P_{\nu})$ in Eq.~(\ref{propagator-7hf}) goes over the $3! = 6$ possible permutations of the indices $\mu_1\mu_2\mu_3(\nu_1\nu_2\nu_3)$.

\subsection{Form factors}

Each hadronic vertex obtained from the Lagrangians given in Sec.~\ref{Sec:Lagrangians} is accompanied by a phenomenological form factor to parametrize the structure of the hadrons and normalize the behavior of the production amplitude. Following Refs.~\cite{Wang:2017,Wang:2018}, for intermediate baryon exchange we take the form factor as
\begin{equation}
f_B(p^2) = \left(\frac{\Lambda_B^4}{\Lambda_B^4+\left(p^2-M_B^2\right)^2}\right)^2,  \label{eq:ff_B}
\end{equation}
where $p$ and $M_B$ denote the four-momentum and the mass of the exchanged baryon $B$, respectively. The cutoff mass $\Lambda_B$ is treated as a fit parameter for each exchanged baryon. For intermediate meson exchange, we take the form factor as
\begin{equation}
f_M(q^2) = \left(\frac{\Lambda_M^2-M_M^2}{\Lambda_M^2-q^2}\right)^2,     \label{eq:ff_M}
\end{equation}
where $q$ represents the four-momentum of the intermediate meson, and $M_M$ and $\Lambda_M$ designate the mass and cutoff mass of the exchanged meson $M$, respectively. In the present work, we use the same cutoff parameter $\Lambda_{K,K^\ast}$ for both $K$ and $K^\ast(892)$ exchanges in the $t$ channel.

Note that the gauge invariance of our photoproduction amplitude is independent of the specific form of the form factors.

\section{Results and discussion}   \label{Sec:results}

As mentioned in Sec.~\ref{Sec:intro}, the reaction $\gamma p\to K^+\Sigma^0(1385)$ was investigated in Refs.~\cite{Yong:2008,hejun:2014} within hadronic models based on effective Lagrangian approaches. In these two works, eight or nine $N$ and $\Delta$ resonances around $2$ GeV (among tens of resonances predicated by a quark model \cite{Capstick:1998}) were considered. The resonance masses were taken from quark model calculations. The resonance hadronic and electromagnetic coupling constants were determined from the resonance decay amplitudes calculated in the quark model \cite{Capstick:1998}. The resonance widths were set to a common value of either $300$ or $500$ MeV. In Ref.~\cite{Yong:2008} it was found that the contributions from the $N$ and $\Delta$ resonances to the cross sections are finite, and in Ref.~\cite{hejun:2014} the contributions from the $N$ and $\Delta$ resonances were even smaller than those in Ref.~\cite{Yong:2008} due to a much larger width being used for all of the resonances.

\begin{table*}[tb]
\caption{\label{Table:chi2} $\chi^2$ per number of data points (ND) for differential cross sections fitted by including one of the resonances below $2200$ MeV. The asterisks below resonance names denote the overall status of these resonances rated by the PDG \cite{Tanabashi:2018}. The numbers in brackets represent the corresponding $\chi^2/$ND for data in the energy range $W\leq 2200$ MeV. Note that $\chi^2/{\text{ND}}=5.1\,(6.4)$ when none of the resonances are taken into account. }
\begin{tabular*}{\textwidth}{@{\extracolsep\fill}lcccccccc}
\hline\hline
$N^\ast$   &  $N(1875){3/2}^-$  & $N(1880){1/2}^+$  &  $N(1895){1/2}^-$  &  $N(1900){3/2}^+$ & $N(2060){5/2}^-$  &  $N(2100){1/2}^+$  &  $N(2120){3/2}^-$  &  $N(2190){7/2}^-$  \\
                 &   $\ast$$\ast$$\ast$       &  $\ast$$\ast$$\ast$        &  $\ast$$\ast$$\ast$$\ast$  &  $\ast$$\ast$$\ast$$\ast$  & $\ast$$\ast$$\ast$        &  $\ast$$\ast$$\ast$         &  $\ast$$\ast$$\ast$        &  $\ast$$\ast$$\ast$$\ast$  \\
$\chi^2/$ND & $2.0\,(2.6)$ & $2.4\,(3.9)$ & $1.6\,(0.9)$ & $2.3\,(2.9)$ & $2.9\,(4.2)$ & $3.1\,(5.3)$  & $3.0\,(4.7)$  & $3.0\,(4.9)$  \\  \hline
$\Delta^\ast$  &  $\Delta(1900){1/2^-}$  &  $\Delta(1905){5/2}^+$ & $\Delta(1910){1/2}^+$ &  $\Delta(1920){3/2}^+$  & $\Delta(1930){5/2}^-$  & $\Delta(1950){7/2}^+$   \\
                      &  $\ast$$\ast$$\ast$               &  $\ast$$\ast$$\ast$$\ast$        &  $\ast$$\ast$$\ast$$\ast$         &  $\ast$$\ast$$\ast$               &  $\ast$$\ast$$\ast$              &  $\ast$$\ast$$\ast$$\ast$  \\
$\chi^2/$ND & $2.1\,(1.5)$ & $2.5\,(3.5)$ & $3.0\,(4.2)$ & $3.4\,(4.8)$ & $2.4\,(1.2)$  &  $3.2\,(3.5)$  \\
\hline\hline
\end{tabular*}
\end{table*}

\begin{figure}[tbp]
\centering\includegraphics[width=\columnwidth]{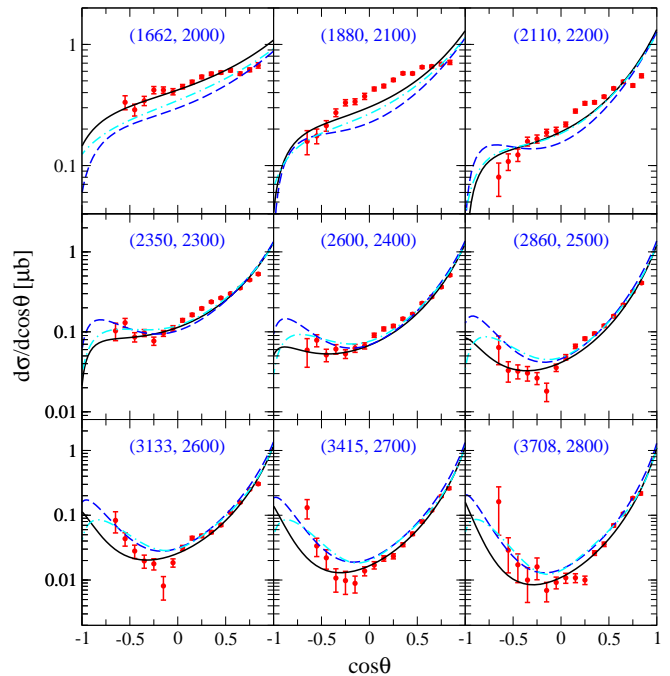}
\caption{Differential cross sections for $\gamma p \to K^+ \Sigma^0(1385)$ as a function of $\cos\theta$. The blue dashed lines represent the results without a resonance contribution. The cyan dash-dotted lines represent the results without a resonance contribution and without the constants from the flavor SU(3) symmetry on the coupling constants in background contributions. The black solid lines represent the results with the exchange of the $N(1875){3/2}^-$ resonance. The scattered symbols denote the data from the CLAS Collaboration \cite{Mori:2013}. The numbers in parentheses denote the centroid value of the photon laboratory incident energy (left number) and the corresponding total center-of-mass energy of the system (right number), in MeV. }
\label{fig:nores_su3_28}
\end{figure}

In the present work, we analyze the available cross-section data for $\gamma p\to K^+\Sigma^0(1385)$ within an effective Lagrangian approach using the tree-level Born approximation, with the purpose of understanding the reaction mechanisms and resonance contents and their associated parameters in this reaction. In addition to the $t$-channel $K$ and $K^\ast(892)$ exchanges, $s$-channel $N$ exchange, $u$-channel $\Lambda$ exchange, and generalized contact term, we introduce a minimum number of $N$ and $\Delta$ resonances in the $s$ channel in constructing the reaction amplitudes to describe the data. We take the PDG values for resonance masses and widths, and treat the products of resonance electromagnetic and hadronic coupling constants as fit parameters due to the lack of experimental information on resonance decays to $K\Sigma(1385)$.

As mentioned above, we introduce as few $N$ and $\Delta$ resonances as possible to describe the data. First, we want to see to what extent we can describe the data if no resonance exchange is taken into account. The corresponding results for differential cross sections are plotted as the blue dashed lines in Fig.~\ref{fig:nores_su3_28}. One sees that the experimental data in the high-energy region can be qualitatively described. However, in the center-of-mass energy region $W\leq 2200$ MeV, the experimental differential cross sections are  significantly underestimated. Even if we further treat the coupling constants $g_{\Lambda NK}$, $g_{\Sigma^\ast NK}$, and $g_{\Sigma^\ast NK^\ast}$ in the nonresonant contributions as fit parameters instead of constraining them by the flavor SU(3) symmetries as expressed in Eqs.~(\ref{g_LamNK})--(\ref{g_SigstNKst}), the resulting differential cross sections (cyan dash-dotted lines in Fig.~\ref{fig:nores_su3_28}) are still in disagreement with the data in the energy region $W\leq 2200$ MeV, indicating the indispensability of the contributions from the $N$ or $\Delta$ resonances. Actually, the nonresonant contributions are mainly constrained by the high-energy data. In particular, the $t$-channel interaction is constrained by the high-energy data at forward angles, and the $u$-channel interaction is constrained by the high-energy data at backward angles. Therefore, within the present model, one cannot reproduce the data in both the high-energy and low-energy regions simultaneously if none of the resonance exchanges are considered.

We then try to introduce one resonance in the $s$ channel in constructing the reaction amplitudes. Since the data in the energy range $W>2200$ MeV can be reproduced by considering the nonresonant contributions as illustrated in Fig.~\ref{fig:nores_su3_28}, we consider only the resonances whose masses are less than $2200$ MeV. We test all of the three-star and four-star $N$ and $\Delta$ resonances one by one in the energy range $1875<W<2200$ MeV with their masses and widths set to their PDG values \cite{Tanabashi:2018}. The corresponding $\chi^2$ per number of data points, $\chi^2/$ND, for the differential cross sections fitted by including one of the three-star or four-star resonances are listed in Table~\ref{Table:chi2}. In this table, the asterisks below resonance names denote the overall status of these resonances rated by the PDG \cite{Tanabashi:2018}. The numbers in brackets represent the corresponding $\chi^2/$ND for data in the energy range $W\leq 2200$ MeV. One sees that by including one of the $N(1895){1/2}^-$, $\Delta(1900){1/2}^-$, and $\Delta(1930){5/2}^-$ resonances, the corresponding $\chi^2/$ND are $1.6$, $2.1$, and $2.4$, respectively, for data in the full energy range considered, and are $0.9$, $1.5$, and $1.2$, respectively, for data in the energy range $W\leq 2200$ MeV. These three fits are treated as acceptable fits and will be discussed in detail later. The fit with the $N(1875){3/2}^-$ resonance results in $\chi^2/{\text{ND}}=2.0$, which is smaller than that of either the fit with the $\Delta(1900){1/2}^-$ resonance, $\chi^2/{\text{ND}}=2.1$, or the fit with the $\Delta(1930){5/2}^-$ resonance, $\chi^2/{\text{ND}}=2.4$, for data in the full energy range considered. However, the fit with the $N(1875){3/2}^-$ resonance has a $\chi^2/{\text{ND}}=2.6$ for data in the energy range $W\leq 2200$ MeV, which is much larger than that of either the fit with the $\Delta(1900){1/2}^-$ resonance, $\chi^2/{\text{ND}}=1.5$, or the fit with the $\Delta(1930){5/2}^-$ resonance, $\chi^2/{\text{ND}}=1.2$. This means that the fit with the $N(1875){3/2}^-$ resonance has a much worse fitting quality for data in the energy range $W\leq 2200$ MeV, as illustrated in Fig.~\ref{fig:nores_su3_28}, where the black solid lines represent the results from this fit. One sees that although the data at most of the energy points can be satisfactorily reproduced, significant discrepancies between the theoretical differential cross sections and the corresponding data are still seen at the energy points $W= 2100$ and $2200$ MeV. Hence, this fit is not acceptable. The fits that include one of the other resonances result in even larger $\chi^2/{\text{ND}}$, indicating even worse fitting qualities, and thus they are also not acceptable fits.

\begin{figure}[tbp]
\centering\includegraphics[width=\columnwidth]{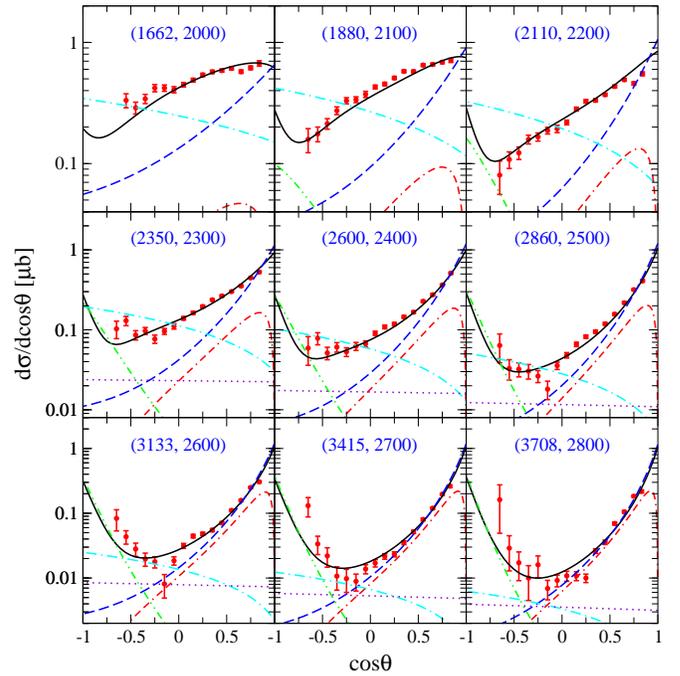}
\caption{Differential cross sections for $\gamma p \to K^+ \Sigma^0(1385)$ as a function of $\cos\theta$ in model I. The black solid lines represent the results from the full calculation. The cyan dash-dotted, blue dashed, green dash-double-dotted, red dot-double-dashed, and violet dotted lines represent the individual contributions from the $s$-channel $N(1895){1/2}^-$ exchange, generalized contact term, $u$-channel $\Lambda$ exchange, $t$-channel $K$ exchange, and $s$-channel $N$ exchange, respectively. The scattered symbols denote the data from the CLAS Collaboration \cite{Mori:2013}. The numbers in parentheses denote the centroid value of the photon laboratory incident energy (left number) and the corresponding total center-of-mass energy of the system (right number), in MeV. }
\label{fig:fit1}
\end{figure}

\begin{figure}[tbp]
\includegraphics[width=\columnwidth]{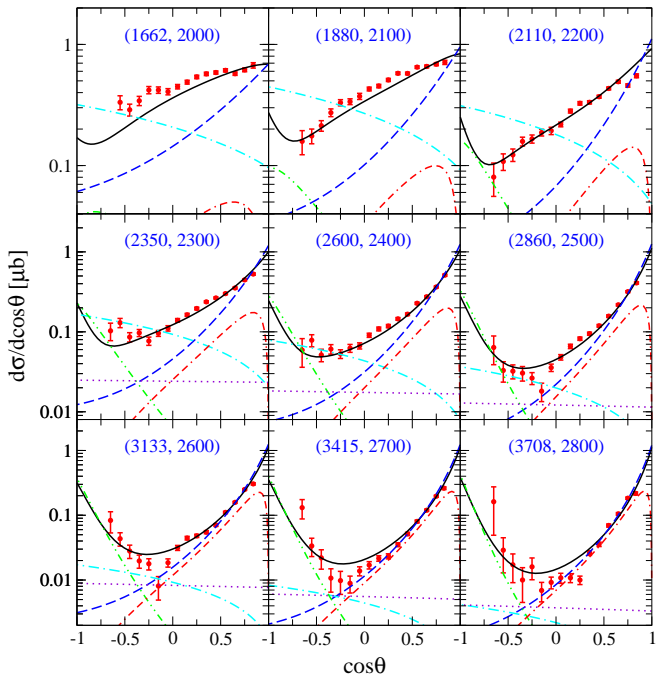}
\caption{Differential cross sections for $\gamma p \to K^+ \Sigma^0(1385)$ as a function of $\cos\theta$ in model II. The notations are the same as in Fig.~\ref{fig:fit1} except that the cyan dash-dotted lines now represent the individual contributions from the $s$-channel $\Delta(1900){1/2}^-$ exchange.}
\label{fig:fit2}
\end{figure}

\begin{figure}[tbp]
\includegraphics[width=\columnwidth]{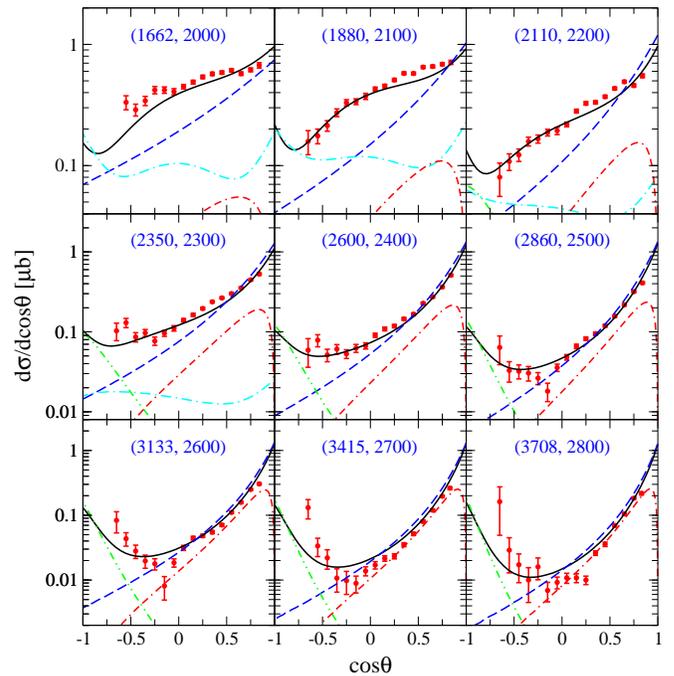}
\caption{Differential cross sections for $\gamma p \to K^+ \Sigma^0(1385)$ as a function of $\cos\theta$ in model III. The notations are the same as in Fig.~\ref{fig:fit1} except that the cyan dash-dotted lines now represent the individual contributions from the $s$-channel $\Delta(1930){5/2}^-$ exchange.}
\label{fig:fit3}
\end{figure}

As discussed above, the most recent available differential cross-section data for $\gamma p\to K^+\Sigma^0(1385)$ from the CLAS Collaboration \cite{Mori:2013} can be satisfactorily described by including one of the $N(1895){1/2}^-$, $\Delta(1900){1/2}^-$, and $\Delta(1930){5/2}^-$ resonances, among which the first one was evaluated by the PDG as a four-star resonance and the other two were evaluated as three-star resonances. We refer to the models including the $N(1895){1/2}^-$, $\Delta(1900){1/2}^-$, and $\Delta(1930){5/2}^-$ resonances as model I, model II, and model III, respectively. The parameters of these three models are listed in Table~\ref{Table:para}, and the corresponding theoretical results for differential cross sections are shown, respectively, in Figs.~\ref{fig:fit1}--\ref{fig:fit3}.

\begin{table}[tb]
\caption{\label{Table:para} Model parameters. The asterisks below resonance names denote the overall status of these resonances evaluated by the PDG \cite{Tanabashi:2018}. The resonance mass $M_R$ and width $\Gamma_R$ are fixed to the values estimated by the PDG, with the numbers in brackets below $M_R$ and $\Gamma_R$ representing the range of the corresponding quantities given by the PDG \cite{Tanabashi:2018}.}
\begin{tabular*}{\columnwidth}{@{\extracolsep\fill}lrrr}
\hline\hline
      &  Model I   &  Model II   &   Model III    \\
\hline
$g^{(1)}_{\Sigma^\ast \Lambda \gamma}/g^{(2)}_{\Sigma^\ast \Lambda \gamma}$  &  $-2.28\pm 0.25$  &  $-1.34\pm 0.31$  &  $-0.60\pm0.20$     \\
$\Lambda_{K,K^\ast}$ [MeV]   &  $924\pm 1$   &   $933\pm 2$   &   $950\pm 1$    \\
$\Lambda_N$ [MeV]   & $1495\pm 11$   &   $1500\pm 10$   &   $800\pm8$    \\
$\Lambda_{\Lambda}$ [MeV]   & $800\pm 10$   &   $838\pm11$   &   $813\pm 9$  \\
\hline
      &  $N(1895){1/2}^-$  &  $\Delta(1900){1/2}^-$   &  $\Delta(1930){5/2}^-$     \\
      &  $\ast$$\ast$$\ast$$\ast$  & $\ast$$\ast$$\ast$  &  $\ast$$\ast$$\ast$  \\
$M_R$ [MeV]  &  ${\bf 1895}$  &  ${\bf 1860}$   &  ${\bf 1950}$  \\
                        &  [1870--1920]  & [1840--1920]  &  [1900--2000]  \\
$\Gamma_R$ [MeV]  &  ${\bf 120}$   &  ${\bf 250}$  &  ${\bf 300}$  \\
                                   &  [80--200]   &  [180--320]  &  [200--400]  \\
$\Lambda_R$ [MeV]   & $1368\pm 8$  &  $1278\pm 10$   &  $943\pm 6$   \\
$g_{RN\gamma}^{(1)}g_{R\Sigma^\ast K}^{(1)}$  & $-3.00\pm 0.06$  &  $3.25\pm 0.08$ & $-0.24\pm 0.06$  \\
$g_{RN\gamma}^{(2)}g_{R\Sigma^\ast K}^{(1)}$  &   &    & $11.59\pm 0.13$  \\
\hline\hline
\end{tabular*}
\end{table}

In Table~\ref{Table:para}, the uncertainties of the values of the fit parameters are estimates arising from the uncertainties (error bars) associated with the fitted experimental differential cross-section data. One sees from Table~\ref{Table:para} that the values of $\Lambda_{K,K^\ast}$, the cutoff parameter for the $t$-channel $K$ and $K^\ast(892)$ exchanges, are close to each other in all three models, indicating similar contributions from $t$-channel $K$ and $K^\ast(892)$ exchanges in these models. The values of $\Lambda_N$, the cutoff parameter for the $s$-channel $N$ exchange, are very close to each other in models I and II, and both are much larger than that in model III, implying much smaller contributions from the $s$-channel $N$ exchange in model III than in models I and II. For the $u$-channel $\Lambda$ exchange, the values of the cutoff parameter $\Lambda_\Lambda$ in models I, II, and III are close to each other, but the values for the coupling constants are not, resulting in different $u$-channel contributions in these three models, as can be seen in Figs.~\ref{fig:fit1}--\ref{fig:fit3}. For $N$ and $\Delta$ resonances, the asterisks below resonance names denote the overall status of these resonances evaluated by the PDG \cite{Tanabashi:2018}. The resonance mass $M_R$ and width $\Gamma_R$ in all three models are not treated as fit parameters, but rather fixed to the corresponding values estimated by the PDG. The numbers in brackets below $M_R$ and $\Gamma_R$ represent the range of the corresponding quantities given by the PDG \cite{Tanabashi:2018}. The resonance cutoff parameter and the products of the resonance hadronic and electromagnetic coupling constants are determined by fits to the differential cross-section data. For the $N(1895){1/2}^-$ resonance, the fitted value of the product of the two coupling constants $g_{RN\gamma}^{(1)}g_{R\Sigma^\ast K}^{(1)}=-3.00$. The PDG suggested a helicity amplitude $A_{1/2}=-0.016$ GeV$^{-1/2}$ for the radiative decay of this resonance, which gives $g_{RN\gamma}^{(1)}=0.083$, resulting in a branching ratio ${\rm Br}[N(1895){1/2}^- \to p\gamma]=0.017\%$. One then gets $g_{R\Sigma^\ast K}^{(1)}=-36.314$, which leads to a branching ratio ${\rm Br}[N(1895){1/2}^- \to K^+\Sigma^0(1385)]=1.34\%$. The $\Delta(1900){1/2}^-$ resonance is below the threshold of $K^+\Sigma^0(1385)$ and we do not discuss its branching ratios. For the $\Delta(1930){5/2}^-$ resonance, the helicity amplitudes from various references quoted by the PDG differ in both amplitude and signs, and moreover, its Breit-Wigner photon decay amplitudes differ from the photon decay amplitudes at the pole in both amplitudes and signs, too. Thus, we will not estimate the branching ratios for this resonance from the fitted product of its electromagnetic and hadronic coupling constants until more reliable values of its photon decay amplitudes become available.

The theoretical results of the differential cross sections for $\gamma p \to K^+ \Sigma^0(1385)$ obtained in models I, II, and III with the parameters listed in Table~\ref{Table:para} are shown in Figs.~\ref{fig:fit1}--\ref{fig:fit3}, respectively. There, the black solid lines represent the results from the full calculation. The blue dashed, green dash-double-dotted, red dot-double-dashed, and violet dotted lines represent the individual contributions from the interaction current (the generalized contact term), $u$-channel $\Lambda$ exchange, $t$-channel $K$ exchange, and $s$-channel $N$ exchange, respectively. The cyan dash-dotted lines in Figs.~\ref{fig:fit1}--\ref{fig:fit3} denote the individual contributions from the $s$-channel $N(1895)1/2^-$, $\Delta(1900)1/2^-$, and $\Delta(1930)5/2^-$ exchanges, respectively. The contributions from the $t$-channel $K^\ast(892)$ exchange are too small to be clearly seen with the scale used, and thus are not plotted. The scattered symbols represent the data from the CLAS Collaboration \cite{Mori:2013}. The numbers in parentheses denote the centroid value of the photon laboratory incident energy (left number) and the corresponding total center-of-mass energy of the system (right number), in MeV.

One sees from Figs.~\ref{fig:fit1}--\ref{fig:fit3} that the overall agreement of our theoretical results with the CLAS differential cross-section data is rather satisfactory in all of the models. In the low-energy region, the differential cross sections are dominated by the generalized contact term and the resonance exchange. In the high-energy region, the differential cross sections at forward angles are dominated by the generalized contact term followed by the $t$-channel $K$ exchange, and the differential cross sections at backward angles are dominated by the $u$-channel $\Lambda$ exchange. In the whole energy region, the contributions from the $t$-channel $K$ exchange in all three models are very similar, which is easily understood since the values of the cutoff parameter $\Lambda_{K}$ are very close to each other in the three models, as listed in Table~\ref{Table:para}. The contributions from the generalized contact term in models I and II are very similar to each other, but both are smaller than those in model III in backward and intermediate angles. This is principally because the contributions of the generalized contact term are relevant to the cutoff parameters $\Lambda_{K}$ and $\Lambda_N$ via the form factors $f_t$ and $f_s$ [cf. Eq.~(\ref{eq:Cmu})], while the values of $\Lambda_{K}$ in all three models are very close to each other, and the values of $\Lambda_N$ in models I and II are almost the same but both are much larger than that in model III. The contributions from the $u$-channel $\Lambda$ exchange are noticeable at backward angles in the high-energy region, and are a little bit bigger in models I and II than in model III. The contributions from the $s$-channel $N$ exchange are visible but small at high energies in models I and II, while they are too small to be plotted in model III due to the much smaller cutoff value of $\Lambda_N$ in model III than in models I and II. The resonance exchange contributes mainly in the low-energy region. One sees that the contributions from the resonance exchange in models I and II are similar to each other. This is mostly because in these two models, the resonances $N(1895)1/2^-$ and $\Delta(1900)1/2^-$ have the same spin and parity quantum numbers, and the difference of the isospin factor can be absorbed into the fit parameter of the coupling constants. In model III, the resonance $\Delta(1930)5/2^-$ exchange contributes noticeably only below $3$ GeV, and overall it is much smaller than the contributions of $N(1895)1/2^-$ and $\Delta(1900)1/2^-$ in models I and II. Of course, the shape of the resonance contribution in model III is quite different than those in models I and II due to the difference of the resonance quantum numbers.

\begin{figure}[tbp]
\includegraphics[width=0.741\columnwidth]{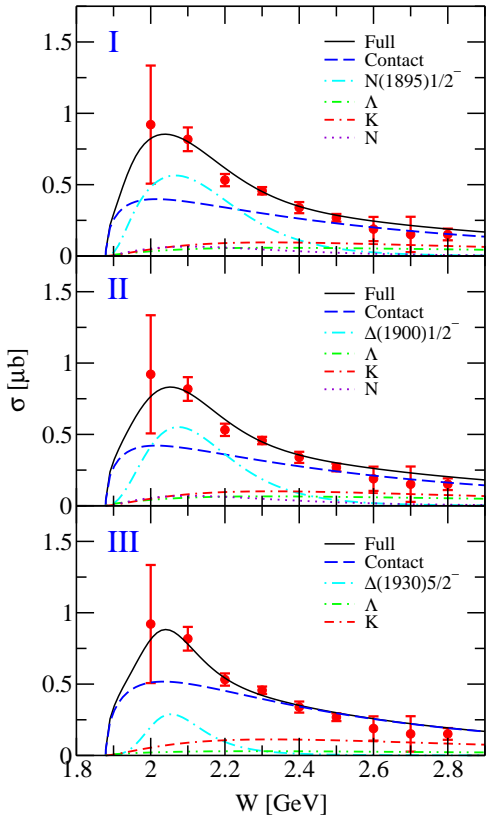}
\caption{Total cross sections with dominant individual contributions for $\gamma p \to K^+ \Sigma^0(1385)$. The panels from top to bottom correspond to the results for modes I--III, as indicated. The data are from CLAS \cite{Mori:2013} but are not included in the fit.}
\label{fig:total_cro_sec}
\end{figure}

Figure~\ref{fig:total_cro_sec} shows our predicted total cross sections (black solid lines) together with individual contributions from the interaction current (blue dashed lines), $s$-channel resonance exchange (cyan dash-dotted lines), $u$-channel $\Lambda$ exchange (green dash-double-dotted lines), $t$-channel $K$ exchange (dot-double-dashed lines), and $s$-channel $N$ exchange (violet dotted lines) obtained by integrating the corresponding results for differential cross sections from models I--III. The contributions from the $t$-channel $K^\ast(892)$ exchange are not plotted since they are too small to be clearly seen with the scale used. Note that the total cross-section data are not included in our fits. One sees from Fig.~\ref{fig:total_cro_sec} that in all three models, our predicted total cross sections are in good agreement with the data over the entire energy region considered. It is seen that in all three models, the generalized contact term has dominant contributions. Actually, as has been discussed in connection with the differential cross section results, the generalized contact term is relevant to the parameters $\Lambda_K$ and $\Lambda_N$ via the $t$-channel and $s$-channel form factors [cf. Eq.~(\ref{eq:Cmu})]. Therefore, in models I and II, the contributions from the generalized contact current are similar, and they both are a little bit smaller than those in model III, since the values of $\Lambda_K$ are similar in models I--III, while the values of $\Lambda_N$ are almost the same in models I and II but both are much larger than that in model III. The contributions from the $t$-channel $K$ exchange are considerable in models I--III, and they are almost the same in all three models due to the similar values of the cutoff parameter $\Lambda_K$. Small but noticeable contributions of the $u$-channel $\Lambda$ exchange to the total cross sections are seen, and these contributions are bigger in models I and II than in model III. The contributions from the $s$-channel $N$ exchange are even smaller than the $u$-channel $\Lambda$ exchange, and in model III they are not plotted as they are too small due to the much smaller cutoff value of $\Lambda_N$ in model III than in models I and II. In all three models, the contributions from the resonances are responsible for the bump structure exhibited by the total cross-section data. It is seen that the resonance exchange provides more important contributions in models I and II than in model III.

\begin{figure}[tb]
\includegraphics[width=\columnwidth]{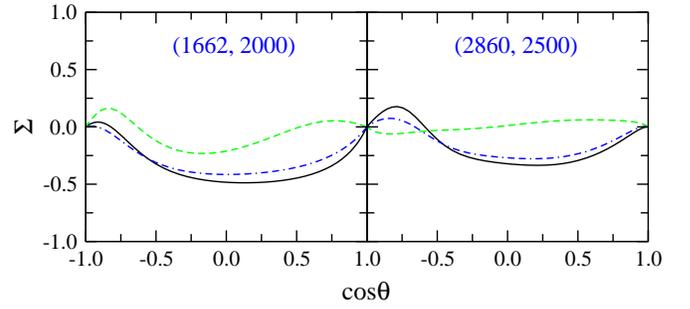}
\caption{Photon beam asymmetries as functions of $\cos\theta$ for $\gamma p \to K^+ \Sigma^0(1385)$. The numbers in parentheses denote the photon laboratory incident energy (left number) and the total center-of-mass energy of the system (right number), in MeV. The black solid, blue double-dash-dotted, and green dashed curves represent the predictions from models I--III, respectively.}
\label{fig:beam_asy}
\end{figure}

\begin{figure}[tb]
\includegraphics[width=\columnwidth]{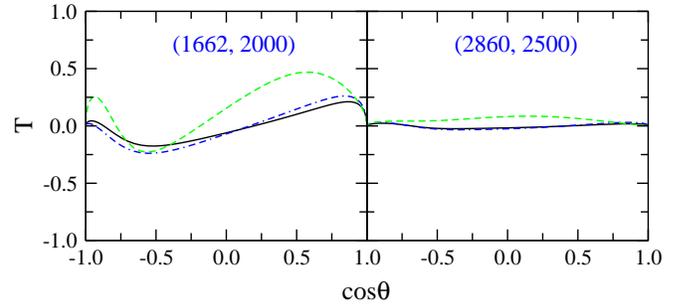}
\caption{Same as in Fig.~\ref{fig:beam_asy} for target nucleon asymmetries.}
\label{fig:target_asy}
\end{figure}

\begin{figure}[tbp]
\includegraphics[width=0.741\columnwidth]{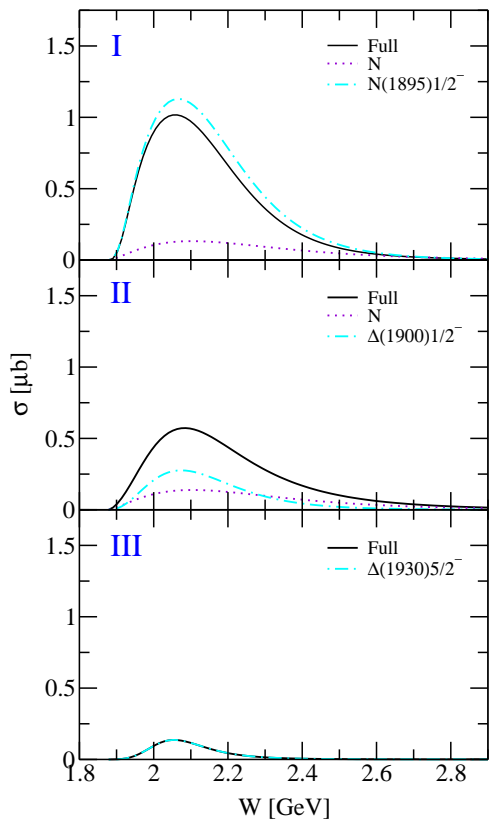}
\caption{Predicated total cross sections with dominant individual contributions for $\gamma p \to K^0 \Sigma^+(1385)$. The panels from top to bottom correspond to the results of models I--III, as indicated. }
\label{fig:total_crosec_2}
\end{figure}

As can be seen in Figs.~\ref{fig:fit1}--\ref{fig:total_cro_sec} and as has been discussed above, models I--III describe the CLAS cross-section data for $\gamma p \to K^+ \Sigma^0(1385)$ quite well with similar fit qualities in the whole energy region considered. Nevertheless, the resonance content of these three models are quite different. It is expected that the spin observables are more sensitive to the dynamical content of various models. In Figs.~\ref{fig:beam_asy}--\ref{fig:target_asy}, we show the predictions of the photon beam asymmetry ($\Sigma$) and target nucleon asymmetry ($T$) from our present models. There, the black solid, blue double-dash-dotted, and green dashed curves represent the predictions from models I--III, respectively. One sees that the $\Sigma$ in model I is similar to that in model II, but both different from that in model III, and similarly for $T$. This means that model III can be distinguished from models I and II by such spin observables, but models I and II are still indistinguishable. This is not a big surprise if one notices that the major difference between models I and II is that the resonance has isospin $1/2$ in model I and $3/2$ in model II, while the resonance isospin factor can be absorbed into the fit parameters of the resonance coupling constants. Therefore, the contributions from all individual terms to differential and total cross sections in model I and model II are almost the same, as can be seen from Figs.~\ref{fig:fit1}--\ref{fig:total_cro_sec}. Given this, one understands that neither $\Sigma$, nor $T$, nor the other spin observables for $\gamma p \to K^+ \Sigma^0(1385)$ can be used to distinguish models I and II. Instead, the cross sections or spin observables for $\gamma p \to K^0 \Sigma^+(1385)$ should be able to distinguish  models I--III due to the different isospin factors.

In Fig.~\ref{fig:total_crosec_2}, we show the predicated total cross sections together with the dominant individual contributions for $\gamma p \to K^0 \Sigma^+(1385)$ in models I--III. Note that there are no free parameters to calculate these results. All of the differences of these contributions for $\gamma p \to K^0 \Sigma^+(1385)$ compared with those for $\gamma p \to K^+ \Sigma^0(1385)$ are due to the different isospin factors for the various interacting terms. In particular, the dominant contributions of the generalized contact term and the considerable contributions of the $t$-channel $K$ exchange in $\gamma p \to K^+ \Sigma^0(1385)$ now vanish in $\gamma p \to K^0 \Sigma^+(1385)$. The contributions from the $N$ resonance exchange in $\gamma p \to K^0 \Sigma^+(1385)$ are double those in $\gamma p \to K^+ \Sigma^0(1385)$, while the contributions from the $\Delta$ resonance exchange in $\gamma p \to K^0 \Sigma^+(1385)$ are half those in $\gamma p \to K^+ \Sigma^0(1385)$. Finally, one sees that the total cross sections for $\gamma p \to K^+ \Sigma^0(1385)$ in models I--III are quite different, unlike the case of $\gamma p \to K^0 \Sigma^+(1385)$ where models I--III give almost the same total cross section values. Therefore, the data on the total cross sections for $\gamma p \to K^0 \Sigma^+(1385)$ could be used to distinguish models I--III. We mention that for the same reason, the other observables of $\gamma p \to K^0 \Sigma^+(1385)$ can also be used to further constrain the theoretical models of $\gamma p \to K^+ \Sigma^0(1385)$.

\section{Summary and conclusion}  \label{sec:summary}

In the present work, we employed an effective Lagrangian approach using the tree-level Born approximation to analyze the most recent differential cross-section data from the CLAS Collaboration for the $\gamma p \to K^+ \Sigma^0(1385)$ reaction. In addition to the $t$-channel $K$ and $K^\ast(892)$ exchanges, $s$-channel $N$ exchange, $u$-channel $\Lambda$ exchange, and generalized contact current, the exchanges of a minimum number of $N$ and $\Delta$ resonances in the $s$-channel were introduced in constructing the reaction amplitudes to describe the data. The $s$-, $u$-, and $t$-channel amplitudes were obtained by evaluating the corresponding Feynman diagrams, and the generalized contact current was constructed in such a way that the full photoproduction amplitudes are fully gauge invariant. It was found that the CLAS differential cross-section data for $\gamma p \to K^+ \Sigma^0(1385)$ \cite{Mori:2013} can be well described by including one of the $N(1895){1/2}^-$, $\Delta(1900){1/2}^-$, and $\Delta(1930){5/2}^-$ resonances, with the resonance mass and width being fixed to their PDG values and the resonance coupling constants being determined by fits to the data. The total cross sections predicated in the theoretical models are in good agreement with the corresponding data.

It was shown that the generalized contact term provides dominant contributions to the differential cross sections of $\gamma p \to K^+ \Sigma^0(1385)$ in the whole energy region considered. The $t$-channel $K$ exchange has important contributions to the differential cross sections at forward angles in the high-energy region, and the $u$-channel $\Lambda$ exchange has considerable contributions to the differential cross sections at backward angles in the high-energy region. The $s$-channel resonance exchange has significant contributions to the differential cross sections in the low-energy region. The total cross sections are dominated by the contributions from the generalized contact term and the $s$-channel resonance exchange, with the latter being responsible for the bump structure exhibited by the CLAS total cross-section data. The $t$-channel $K$ exchange has noticeable but small contributions to the total cross sections. The $u$-channel $\Lambda$ exchange followed by the $s$-channel $N$ exchange has even smaller contributions to the total cross sections than the $t$-channel $K$ exchange.

The predictions of the photon beam asymmetry ($\Sigma$) and target nucleon asymmetry ($T$) from our theoretical models were also presented for the $\gamma p\to K^+\Sigma^0(1385)$ reaction. Their shapes in models I and II are similar, and both are different from those in model III. The predications of the total cross sections for the $\gamma p\to K^0\Sigma^+(1385)$ reaction were also given, which were shown to be quite different in various theoretical models, and are expected to further constrain the theoretical models for $\gamma p\to K^+\Sigma^0(1385)$, leading to a better understanding of the reaction mechanisms, resonance contents, and associated parameters in this reaction.

\begin{acknowledgments}
This work is partially supported by the National Natural Science Foundation of China under Grants No.~11475181 and No.~11635009, and the Key Research Program of Frontier Sciences of Chinese Academy of Sciences under Grant No. Y7292610K1.
\end{acknowledgments}


\begin{thebibliography}{99}
%
\bibitem{Isgur:1978}
N. Isgur and G. Karl, Phys. Rev. D {\bf 18}, 4187 (1978).
%
\bibitem{Capstick:1986}
S. Capstick and N. Isgur, Phys. Rev. D {\bf 34}, 2809 (1986).
%
\bibitem{Loring:2001}
U. L\"{o}ring, B. C. Metsch, and H. R. Petry, Eur. Phys. J. A {\bf 10}, 395 (2001).
%
\bibitem{CBCG67}
J. H. R. Crouch {\it et al.} (Cambridge Bubble Chamber Group), Phys. Rev. {\bf 156}, 1426 (1967).
%
\bibitem{DBCG67}
R. Erbe {\it et al.} (DESY Bubble Chamber Group), Nuovo Cimento A {\bf 49}, 504 (1967).
%
\bibitem{ABBH69}
R. Erbe {\it et al.} (ABBHHM Collaboration), Phys. Rev. {\bf 188}, 2060 (1969).
%
\bibitem{Mori:2013}
K. Moriya {\it et al.}  (CLAS Collaboration), Phys. Rev. C {\bf 88}, 045201 (2013).
%
\bibitem{Yong:2008}
Y. Oh, C. M. Ko, and K. Nakayama, Phys. Rev. C {\bf 77}, 045204 (2008).
%
\bibitem{Capstick:1998}
S. Capstick and W. Roberts, Phys. Rev. D {\bf 58}, 074011 (1998).
%
\bibitem{hejun:2014}
J. He, Phys. Rev. C {\bf 89}, 055204 (2014).
%
\bibitem{Yu:2017}
Byung-Geel Yu and Kook-Jin Kong, Phys. Rev. C {\bf 95}, 065210 (2017).
%
\bibitem{Tanabashi:2018}
M. Tanabashi {\it et al.} (Particle Data Group), Phys. Phys. D {\bf 98}, 030001 (2018).
%
\bibitem{Haberzettl:1997}
H. Haberzettl, Phys. Rev. C {\bf 56}, 2041 (1997).
%
\bibitem{Haberzettl:2006}
H. Haberzettl, K. Nakayama, and S. Krewald, Phys. Rev. C {\bf 74}, 045202 (2006).
%
\bibitem{Huang:2012}
F. Huang, M. D\"{o}ring, H. Haberzettl, J. Haidenbauer, C. Hanhart, S. Krewald, U.-G. Mei{\ss}ner, and K. Nakayama, Phys. Rev. C {\bf 85}, 054003 (2012).
%
\bibitem{Huang:2013}
F. Huang, H. Haberzettl, and K. Nakayama, Phys. Rev. C {\bf 87}, 054004 (2013).
%
\bibitem{Garcilazo:1993}
H. Garcilazo and E. Moya de Guerra, Nucl. Phys. {\bf A562}, 521 (1993).
%
\bibitem{Swart:1963}
J. J. Swart, Rev. Mod. Phys. {\bf 35}, 916 (1963).
%
\bibitem{Ronchen:2013}
D. R\"{o}nchen, M. D\"{o}ring, F. Huang, H. Haberzettl, J. Haidenbauer, C. Hanhart, S. Krewald, U.-G. Mei{\ss}ner, and K. Nakayama, Eur. Phys. J. A {\bf 49}, 44 (2013).
%
\bibitem{Kim:2014}
S. H. Kim, A. Hosaka, and H. C. Kim, Phys. Rev. D {\bf 90}, 014021 (2014).
%
\bibitem{Wang:2017}
A. C. Wang, W. L. Wang, F. Huang, H. Haberzettl, and K. Nakayama, Phys. Rev. C {\bf 96}, 035206 (2017).
%
\bibitem{Wang:2018}
A. C. Wang, W. L. Wang, and F. Huang, Phys. Rev. C {\bf 98}, 045209 (2018).
%
\bibitem{Kim:2011}
S. H. Kim, S. Nam, Y. Oh, and H. C. Kim, Phys. Rev. D {\bf 84}, 114023 (2011).
%
\bibitem{Behrends:1957}
R. E. Behrends and C. Fronsdal, Phys. Rev. {\bf 106}, 345 (1957).
%
\bibitem{Fronsdal:1958}
C. Fronsdal, Suppl. Nuovo Cimento {\bf 9}, 416 (1958).
%
\bibitem{Zhu:1999}
J. J. Zhu and M. L. Yan, arXiv:hep-ph/9903349.
%
\end{thebibliography}
\end{document}